\pdfoutput=1
\documentclass[twocolumn,aps,showpacs,preprintnumbers,amsmath,amssymb]{revtex4}

\usepackage{graphicx}
\usepackage{dcolumn}
\usepackage{bm}

\usepackage[english]{babel}
\usepackage[T1]{fontenc}
\usepackage[ansinew]{inputenc} 

\begin{document}


\title{Ion traps with enhanced optical and physical access}

\author{Robert Maiwald}
\email{Robert.Maiwald@physik.uni-erlangen.de}
 \affiliation{Institute of Optics, Information and Photonics, University Erlangen-Nuremberg, 91058 Erlangen, Germany}
\author{Dietrich Leibfried}
 \email{dil@boulder.nist.gov}
\affiliation{Time and Frequency Division, National Institute of Standards and Technology,
Boulder, CO 80305, USA}

\author{Joe Britton}
\affiliation{Time and Frequency Division, National Institute of Standards and Technology,
Boulder, CO 80305, USA}
\author{J. C. Bergquist}
\affiliation{Time and Frequency Division, National Institute of Standards and Technology,
Boulder, CO 80305, USA}
\author{Gerd Leuchs}
\affiliation{Institute of Optics, Information and Photonics, University Erlangen-Nuremberg, 91058 Erlangen, Germany}
\author{D. J. Wineland}
\affiliation{Time and Frequency Division, National Institute of Standards and Technology,
Boulder, CO 80305, USA}

\date{19th March 2009}

\begin{abstract}
Small, controllable, highly accessible quantum systems can serve as probes at the single quantum level to study multiple physical effects, for example in quantum optics or for electric and magnetic field sensing. The applicability of trapped atomic ions as probes is highly dependent on the measurement situation at hand and thus calls for specialized traps. Previous approaches for ion traps with enhanced optical access included traps consisting of a single ring electrode \cite{YU91, SCH93} or two opposing endcap electrodes \cite{SCH93, DES06}. Other possibilities are planar trap geometries, which have been investigated for Penning traps \cite{STA04,CAS07} and rf-trap arrays \cite{CHI05,SEI06,PEA06}. By not having the electrodes lie in a common plane the optical access in the latter cases can be substantially increased. Here, we discuss the fabrication and experimental characterization of a novel radio-frequency (rf) ion trap geometry. It has a relatively simple structure and provides largely unrestricted optical and physical access to the ion, of up to 96\% of the total 4$\pi$ solid angle in one of the three traps tested.  We also discuss potential applications in quantum optics and field sensing.  As a force sensor, we estimate sensitivity to forces smaller than 1 yN~Hz$^{-1/2}$.
\end{abstract}

\maketitle

\section{\label{sec:Model}Basic geometry}

\begin{figure}
\includegraphics[width=8cm]{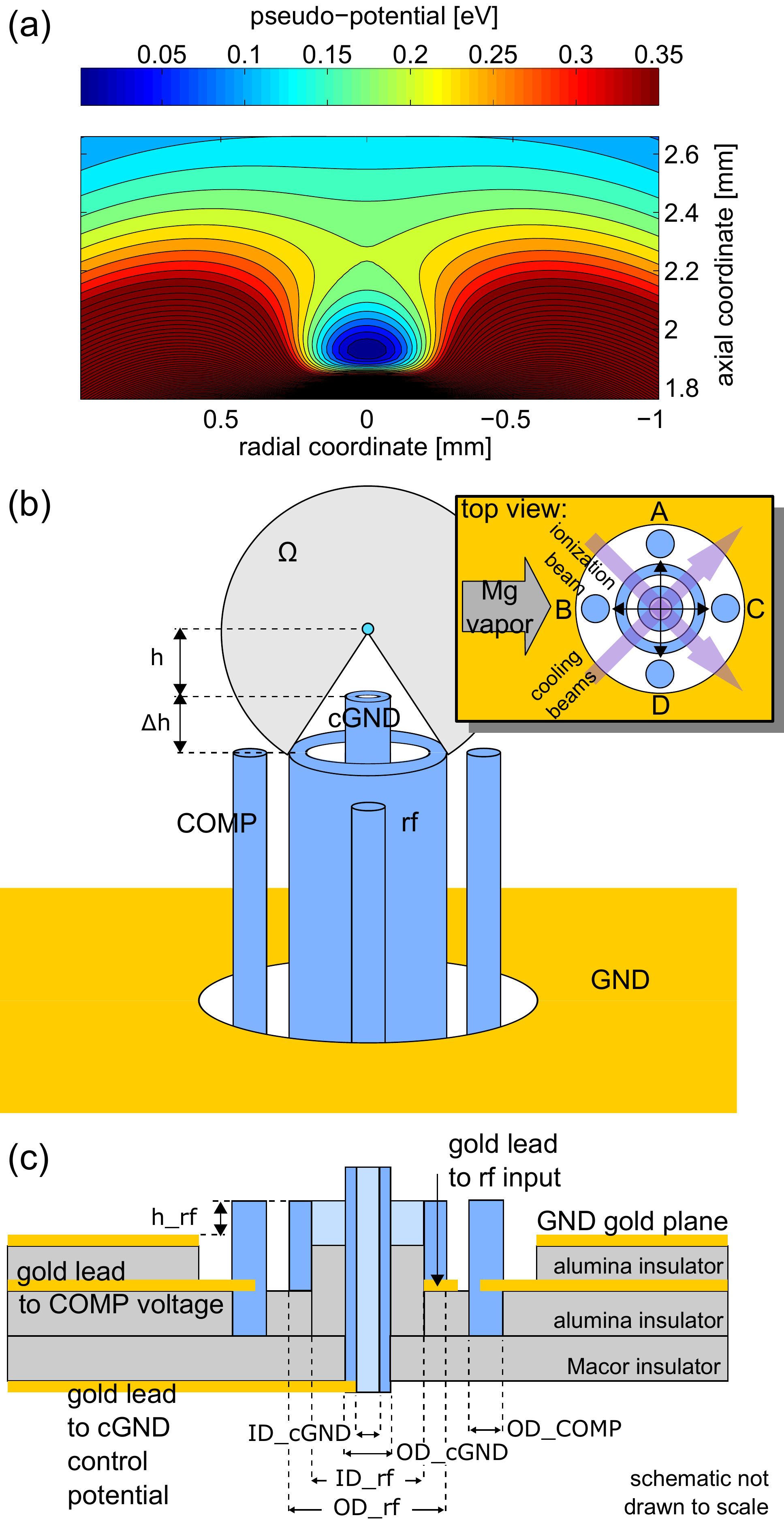}
\caption{\label{fig:TrapGeom} (a) Simulations of the trap rf pseudopotential indicate smaller trap potential depths and binding frequencies compared to those of more
traditional trap designs of similar size, rf-drive frequency and
amplitude. Shown is an example calculation of the pseudo-potential (in eV) for electrode configuration \#3 using the corresponding parameters given in Table \ref{tab:trapData} and $^{24}$Mg$^+$ properties, while neglecting the compensation electrodes. Isoline
separation is 25 meV and the axial coordinate is measured from the grounded plane. (b) Placement of the
electrodes. The central ground electrode (cGND) is surrounded by the
rf electrode (rf) and the grounded plane (GND). These electrodes provide the primary trapping potential. In addition, four symmetrically placed compensation electrodes (COMP) provide fine adjustments to the overall potential. The distance $h$ between the ion and the center electrode varies
with $\Delta h$, the height difference between the center electrode and
the rf electrode. The accessible solid angle $\Omega$ is also illustrated. The inset shows the position of the trap electrodes with respect to the laser beams for cooling and ionization and the direction of the principal radial axes of the trap. The direction from which the neutral Mg vapor enters the trapping region is indicated. (c) The layered assembly showing insulating planes with conducting gold leads and laser-cut holes to house the electrodes.}
\end{figure}

The basic electrode geometry is shown in Fig. {\ref{fig:TrapGeom}} and is formed by two concentric cylinders over a ground
plane. The design provides straightforward indexing and assembly of the trap electrodes, with large solid angle access to the ion. Four
additional electrodes were placed on a circle between the grounded plane and the rf electrode to break the rotational symmetry of the rf pseudopotential about the vertical axis and to compensate for stray electric fields in order to minimize ion rf-micromotion in the trap \cite{BER98}.

Three different traps were built adjacent to each other on the same test set-up. These traps range from a conservative design with larger trap depth, higher motional frequencies and a smaller accessible solid angle, to a weaker trap with greater optical access. This change in properties is achieved by varying the protrusion height $\Delta h$ of the central grounded electrode with respect to the rf electrode (Table \ref{tab:trapData}).

The degeneracy of motional frequencies in the radial direction was lifted by applying potentials on the order of 0.1 to 1\;V to the compensation electrodes $A-D$. This created a static quadrupole field defining the principal radial axes of the trap along the lines connecting compensation electrode A with D and B with C. Thus the axes were oriented at angles of about 45$^\circ$ relative to the two cooling beams (Fig. {\ref{fig:TrapGeom}}b). In addition, the entire trap assembly was tilted by about 7.5$^\circ$ with respect to the direction defined by the
laser beams, ensuring that the vertical axis of the traps was not orthogonal to the wavevectors of the cooling beams. In this way all three normal modes of the ion were sufficiently Doppler cooled by a single laser beam.

\section{\label{sec:Construct} Trap construction}

The trap electrodes are made of stainless steel rods and hypodermic tubing. The dimensions indicated in Fig. {\ref{fig:TrapGeom}} are as follows: $\rm{OD\_rf}=710\;\mu m$,
$\rm{ID\_rf}=535\;\mu m$, $\rm{OD\_cGND}=205\;\mu m$,
$\rm{ID\_cGND}=100\;\mu m$, $\rm{OD\_COMP}=150\;\mu m$, $\rm{h\_rf}=1110\;\mu m$. Structural integrity and insulation is provided by a combination of alumina and Macor spacers.
Electrical connections were made by resistive
welding of gold ribbons to the trap electrodes and to gold
traces (thickness 5-10\;$\mu$m) that were silk-screened onto alumina. The outer ground plane was also formed by this silk-screening process. All insulating
surfaces have been recessed or positioned to prevent a direct line of sight to the ion position. This suppresses distortions
of the trapping field caused by charged insulating surfaces. The potentials applied to the individual DC electrodes (central ground and
compensation electrodes $A-D$) are passively filtered by \textit{RC} low-pass filters with $R=1\;\rm{k\Omega}$ and $C=2\;\rm{nF}$. The
components are mounted on a connection board inside the vacuum close to the trap.

\begin{figure}[b]
\includegraphics[width=8.5cm]{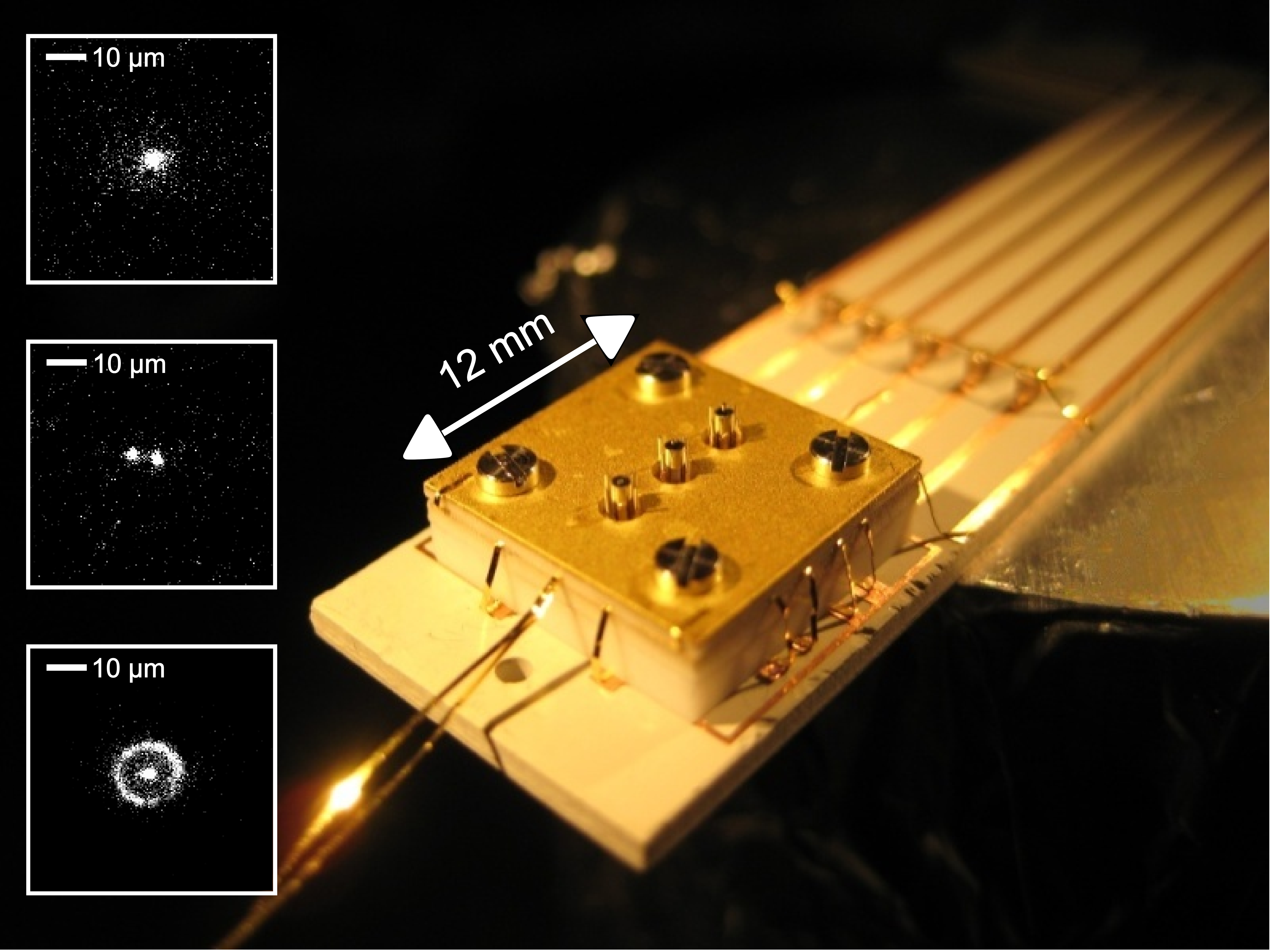}
\caption{\label{fig:TrapPhoto} The three test traps with
differing rf electrode protrusion height $\Delta h$ (increasing from front to back) of the
central electrode can be seen. Visible in the upper right-hand side of the figure is also the ceramic board with copper traces, that accommodates surface mounted components constituting low-pass filters for the DC electrodes. Two parallel gold wires connecting the rf-electrode can be seen in the lower left-hand side of the figure. The insets show different experimentally observed ion configurations (from top to bottom): a single trapped ion, two ions revealing the orientation of the weakest trap axis, and an ion crystal in an approximately cylindrical potential that allows the outer ions to rotate about the trap vertical axis.}
\end{figure}

The assembled trap package is attached to the connection board (Fig.
{\ref{fig:TrapPhoto}}) which is in turn is mounted inside a copper
tube that forms part of a rf resonator that supplies the rf potential. The traps, Mg ovens, and copper tube reside inside a quartz envelope with
extrusions and flat windows for laser beams and for imaging of the
trapped ions. The vacuum system is completed by a 20 liter/s
ion getter pump, a Ti-sublimation pump and an ionization pressure
gauge. The system was pumped to 5$\times$10$^{-5}$\;Pa and
baked-out at 210\;$^{\circ}$C for 8 days. After cooling the system
to room temperature and several applications of Ti-sublimation, a base
pressure of about 3$\times$10$^{-9}$\;Pa (at the gauge) was achieved.

Helical resonators were driven by a low-noise signal
generator to generate the rf trap potentials. They had loaded $Q$-factors
ranging from 300 to 430 and were driven with input powers in the
range of 22 to 35 dBm. This resulted in rf amplitudes (inferred from simulations and the resulting trap frequencies) in the range of 290 to 460\;V.

\section{\label{sec:Experiment} Experimental setup and trap characterization}

A $^{24}$Mg oven produced a vapor of neutral atoms that was directed at the traps in the horizontal direction of Fig.
\ref{fig:TrapGeom}b. To load ions into the traps,
the neutral atoms inside the confinement region were then photoionized. All experiments were performed at a magnetic field of $B\simeq$ 0 T.
Photoionization and Doppler laser cooling were achieved by crossing three laser
beams in the confinement region of a selected trap. The photo-ionization beam ($\lambda=285$\;nm, 2 to 4 mW, 50 $\mu$m waist) was resonant with
the $3s^2$ $^1S_0$ $\leftrightarrow$ $3s3p$ $^1P_1$ transition in
neutral Mg \cite{EPS07}. A second photon either at 285 nm or at 280 nm (below)
promotes the electron from the $3s3p$ $^1P_1$ state to the continuum,
producing a $^{24}$Mg$^+$ ion in the confinement region. For
cooling, a second laser beam ($\lambda=280$\;nm, 1 mW, 30 $\mu$m waist) was tuned about 400 MHz
below the $3s^2$ $^2S_{1/2}$ $\leftrightarrow$ $3s3p$ $^2P_{1/2}$
transition of $^{24}$Mg$^{+}$ for initial cooling. A third, much
weaker beam ($\lambda=280$\;nm, 2-10 $\mu$W), with intensity below saturation and
tuned below the $3s^2$ $^2S_{1/2}$ $\leftrightarrow$ $3s3p$ $^2P_{1/2}$ transition by half of the natural linewidth ($\simeq$ 20 MHz), subsequently cooled the ions to near the Doppler limit. At that point photon scattering and cooling was dominated by the closely detuned beam. In the
case of sudden ion heating, for example due to a collision with
background gas, the 400 MHz detuned beam could efficiently recool the ion
to a point where the near-resonant beam would take over.

The ion was detected by collecting fluorescence with a high
numerical aperture (NA $\simeq$ 0.4) objective located a distance of 5 cm
above the ion, and either imaged onto an electron multiplying charge coupled imaging device (EMCCD) or photomultiplier tube.

To compensate rf-micromotion due to electric stray fields, we varied the potential on all four compensation electrodes and the central
cylinder while changing the power level of the rf drive. Compensation was achieved when the ion position remained stationary
for different rf drive amplitudes \cite{BER98}. We could also minimize micromotion along the direction of the cooling beam by adjusting compensation potentials and maximizing the scattering signal with the cooling beam tuned close to resonance. By lowering the rf drive
after loading we determined a minimum rf power for ``stable'' trapping which seemed to be limited only by background gas collisions.
The trap of configuration \#3 was able to trap reliably after the well depth was lowered to less than 1\;meV. Since this depth is significantly below room temperature, ions can be lost by a single Langevin (charge-dipole) collision with a background gas atom. At the base pressure of about 3$\times$10$^{-9}$\;Pa reached in our system, the mean ion lifetime with laser cooling was around 30 s in this particular case.

In all three traps, we observed ion lifetimes in excess of three hours under continuous laser cooling. Without cooling light, lifetimes were
greater than 10 s. The secular frequencies of ion motion along all three axes were determined by applying additional sinusoidal drive potentials to the compensation electrodes or the central electrode. If the drive frequency is resonant with a secular frequency, ion motion is excited to large amplitudes, causing a sharp drop of laser induced fluorescence.

Operating parameters for each of the three traps are summarized in Table \ref{tab:trapData} together with observed trap
frequencies, rf-drive voltage and trap depth as inferred from measured trap frequencies and numerical simulations for each
trap. Accessible solid angle ignores the compensation electrodes and outer ground plane. This seems reasonable because in future implementations, these electrodes could be made smaller and/or recessed below the rf electrode. This has not been done so far to facilitate construction of the basic electrode structure.

\begin{table}[b]
\caption{\label{tab:trapData} These typical operating parameters were derived from measurements on single trapped $^{24}$Mg${^+}$ ions. The main difference between the three traps is the
protrusion length $\Delta h$ of the center electrode beyond the rf
electrode (Fig. \ref{fig:TrapGeom}b). Due to this variation, different trap frequencies and
accessible solid angles are obtained. Some parameters (denoted by
$^{\blacktriangle}$) are difficult to measure directly and were
inferred from the measured trap frequencies and the numerical
simulation of the trapping potentials. Radial $\overline{AD}$ ($\overline{BC}$) indicates a normal mode direction along the line connecting compensation electrodes $A$ and $D$ ($B$ and $C$), see also Fig. \ref{fig:TrapGeom}b. The observed distances $h$ were 10-20~\% smaller than predicted by simulations.  This seems reasonable since the simulations neglected the rf-grounded compensation electrodes, which if included, would reduce the predicted values of $h$.}
\begin{ruledtabular}
\begin{tabular}{l|ccc|c}
electrode configuration & trap \#1 & trap \#2 & trap \#3 & unit \\
\hline
protrusion height $\Delta h$ & 0 & 250 & 500 & $\mu$m \\
\hline
rf drive voltage $U$\,$^{\blacktriangle}$ & 290 & 460 & 400 & V \\
\hline
rf drive frequency $\Omega_{\rm{rf}}/(2 \pi)$ & 80.15 & 31.94 & 11.85 & MHz \\
\hline
trap frequencies: & & & & \\
- axial & 1.8 & 2.2 & 2.1 & MHz \\
- radial $\overline{AD}$ & 0.951 & 1.268 & 1.064 & MHz \\
- radial $\overline{BC}$ & 0.907 & 1.233 & 1.007 & MHz \\
\hline
trap depth\,$^{\blacktriangle}$ & 71 & 178 & 195 & meV \\
\hline
observed distance $h$ & 168 & 244 & 290 & $\mu$m \\
\hline
accessible solid angle $\Omega/(4\pi)$ & 71\% & 91\% & 96\% & \\
\end{tabular}
\end{ruledtabular}
\end{table}

To compensate micromotion due to electric stray fields, we varied
the potential on all four compensation electrodes and the central
tube while changing the power level of the rf drive. Compensation was achieved when the ion position remained stationary
for different rf drive amplitudes \cite{BER98}. In addition, we could minimize
micromotion along the direction of the cooling beam by maximizing
the scattering signal close to resonance. By lowering the rf drive
after loading we determined a minimum ``stable'' trapping rf power
level which seemed to be limited only by background gas collisions.
A trap loaded in configuration \#3 was able to trap reliably
after the trap well depth was lowered to less than 1\;meV. For a well depth below room temperature ions can be lost by a single Langevin (charge-dipole) collision with a background gas atom. At the base pressure reached in our system, the mean ion lifetime with laser cooling on was around 30 s in this particular case.

\begin{figure}[b]
\includegraphics[width=5cm]{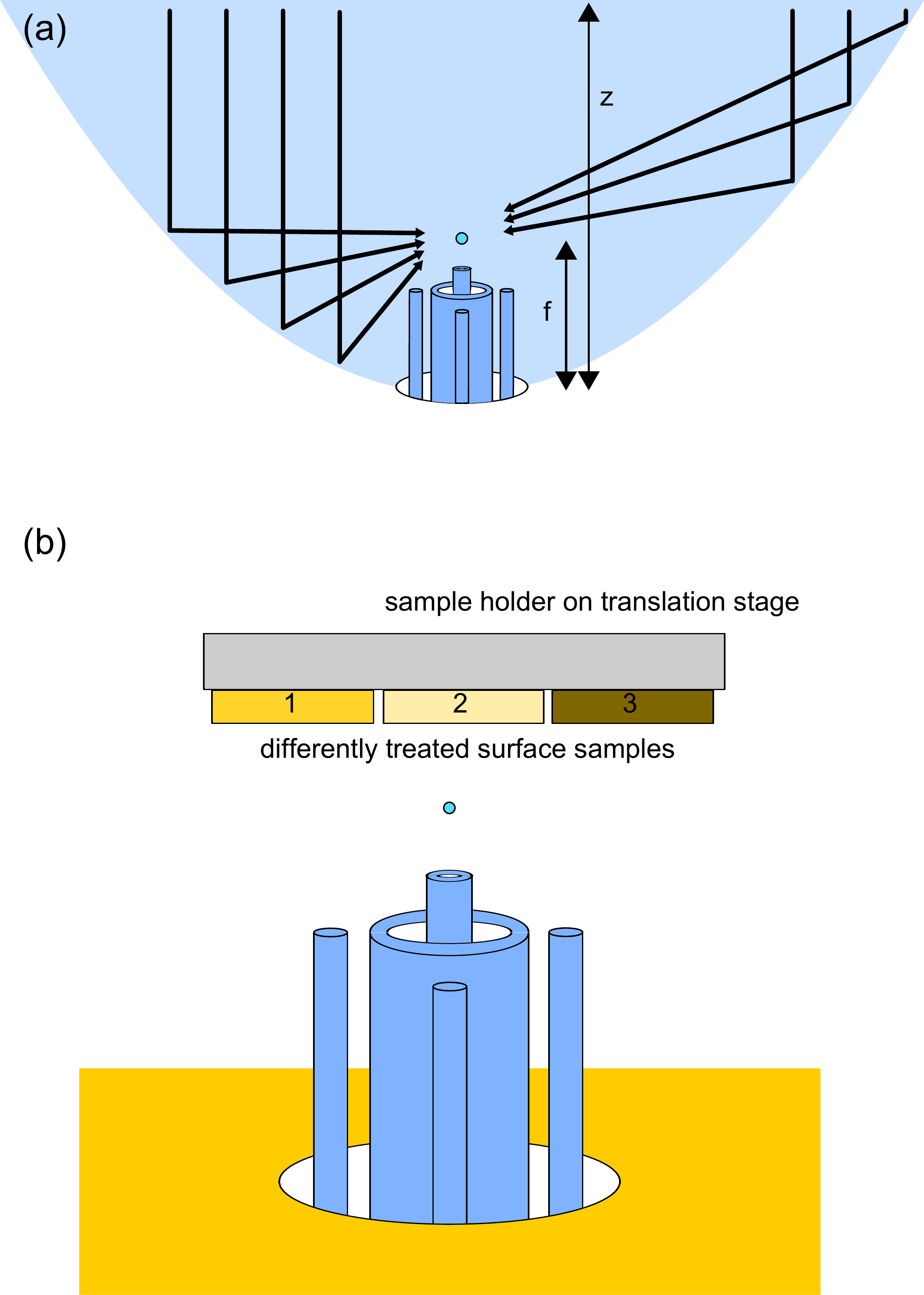}
\caption{\label{fig:Apps} (a) Placement of the ion in the focus $f$ of a parabolic mirror with depth $z$ to maximize photon-ion coupling. (b) Scanning of different surfaces with the ion as a sensitive probe.}
\end{figure}

\section{\label{sec:QuantOpt} Large solid angle photon absorption and emission}

The compact design permits placing the trap inside a metallic parabolic mirror, which can also serve as an rf ground electrode (Fig. {\ref{fig:Apps}}a). By moving the trap structure, an ion can be placed at the focal point of the mirror for efficient fluorescence collection. Moreover, parallel light beams directed along the mirror optical axis are focused directly onto the ion. By appropriately shaping the
transverse mode pattern of the incident light \cite{QUA00, LIN07} the field at the focus can produce a linear dipole excitation
pattern aligned with the axis of symmetry of the parabolic mirror, which could lead to very efficient photon-ion coupling \cite{SON07}. With electrode configuration \#3 and a parabolic mirror with a depth-to-focal-length ratio of 6:1 (an aspect ratio that is feasible to manufacture) the solid angle intercepted by the parabolic mirror is 81~\% of 4$\pi$. For a linear electric dipole aligned along the mirror axis, this geometry would lead to a collection efficiency of 94~\%. Conversely, light sent onto the ion in a dipolar pattern would provide a near perfect atom-to-photon coupling to a linear dipole transition, possibly working down to the single-photon level \cite{STO08, PIN08}. To this end one could use ions with even-numbered charge which allow for J=0 to J=1 transitions \cite{SON07}.

This scheme might also provide a significant improvement in the efficiency of remote entanglement of trapped ions as described in Refs. \cite{MOE07,GER08,OLM09}. As an example, the entanglement scheme used in \cite{OLM09} is based on filtering the emission of the atom on F'=0,1 to F=0,1 transitions to only overlap photons on a beamsplitter resulting from $\pi$-transitions. Limitations of the experiment in \cite{OLM09} were caused by restrictions in collection solid angle (0.02~\% of 4$\pi$) as well as coupling of the emission pattern that was imaged through a multi-element lens into a single mode fiber (efficiency 20~\%). A parabolic collection mirror could potentially improve the situation because ideally it transforms the orthogonal mode patterns of photons emitted on $\pi$ or $\sigma$-transitions while preserving their orthogonality \cite{SON08}. Orthogonality is also preserved in good approximation by the remaining elements of the mode shaping envisioned in \cite{QUA00, LIN07}. Therefore it might be possible to not only couple photons from $\pi$-transitions with near unit efficiency to a single mode fiber, but the mode converter will also act as a filter blocking out undesired photons from $\sigma$-transitions. Ideally one would expect to boost the production rate of entangled pairs by more than $5\times 10^4$ over the values reported in \cite{OLM09}. Efficient coupling could also be obtained through a resonant interaction of an ion with a cavity \cite{MUN02,KEL04}. In this method, to achieve a coupling efficiency of 90~\%, a minimum cooperativity of 4.5 is required. Currently, this is difficult to achieve for many ions that have ultraviolet transition wavelengths and additional complications may be encountered with mirror charging \cite{MUN02,KEL04}.

\section{\label{sec:SurChar} Ion trap surface sensor}

The open geometry of such traps also suggests applications as a probe of fields near a surface. Surfaces of interest could be brought close to the ion, as in Fig. {\ref{fig:Apps}}b. By either scanning the ion trap over the surface or translating the surface itself, one could map out the electromagnetic or force fields in proximity of the surface. Modeling the ion trap in the presence of a ground plane located horizontally above the ion in Fig. {\ref{fig:TrapGeom}} indicates  that a stable quadrupole minimum is retained until the distance to the surface is approximately equal to the distance of the ion to the center electrode. In the traps described here, this would limit the ion to distances of about 170 $\mu$m from the surface, a limit that could be reduced by miniaturizing the trap. The ion serves as a stylus probe tip that is extremely sensitive to forces oscillating at its motional frequencies. These frequencies can be tuned over at least two decades from approximately $\omega/(2 \pi)$ = 100 kHz to 10 MHz. In addition, information on the force field direction can be extracted by utilizing all three nondegenerate modes of motion of the ion. With the same apparatus, static and time-dependent magnetic fields can be measured by observing first and second order shifts of the ion on narrow internal transitions.

To estimate the sensitivity to oscillating force fields we assume that the ion, initially cooled to its motional ground state, is driven in resonance with a motional mode for a duration $t=1/\Delta_b$, where $\Delta_b$ is the approximate measurement bandwidth. The amplitude $\alpha$ of the coherent state grows as
\cite{CAR65}
\begin{equation}
\alpha = \frac{F z_0}{2 \hbar} t,
\end{equation}
where $F$ is the amplitude of the driving force, $z_0 = \sqrt{\hbar/(2 m \omega)}$ is the size of the ion's harmonic
oscillator ground state wave function, $m$ is the ion's mass, and $\hbar$ is Planck's constant divided by 2$\pi$. To yield a detectable
signal, this coherent excitation has to be comparable to the excitation due to motional heating that grows according to $\langle n_n\rangle=\langle \dot{n}\rangle t$ where $\langle \dot{n}\rangle$ is the ion's heating rate \cite{TUR00,LAB08}. Heating rates in the range of 0.2 quanta/ms to 2 quanta/ms have been observed in traps with similar electrode-to-ion distances for  a motional frequency  $\omega/(2 \pi) \simeq 1$~MHz \cite{DES04,EPS07}. Since the  geometry discussed here minimizes the amount of material close to the ion a heating rate of 1 quantum per millisecond should be a realistic estimate. (For small cryogenic traps, heating rates on the order of 1 quantum per second have been observed \cite{LAB08}.) Assuming a signal-to-noise ratio of one, we require $\langle n_n\rangle =\langle n_c\rangle =|\alpha|^2$ and therefore
\begin{equation}
\frac{F}{\sqrt{\Delta_b}} \simeq \sqrt{\langle \dot{n}\rangle}\frac{2 \hbar}{z_0}.
\end{equation}
 For $^{24}$Mg$^+$, $\langle \dot{n}\rangle$ = 1/ms and $\omega/(2 \pi)$ = 1 MHz, this implies a force sensitivity of 0.46 {\rm yN}/$\sqrt{{\rm Hz}}$ (1 yN = 1 yocto-Newton = $10^{-24}$N), several orders of magnitude below the smallest forces detectable with atomic force microscopes or micro-mechanical cantilevers \cite{MAM01}.
This force sensitivity corresponds to an electric field sensitivity of 2.9 $(\mu$V/m)/$\sqrt{{\rm Hz}}$. For a cryogenic ion trap where $\langle \dot{n}\rangle$~=~1/s the sensitivity would be increased by approximately a factor of 30.

To detect the magnetic field at the ion's position we may excite a narrow transition (for example a
hyperfine transition) with well known field dependence. The method would be limited by quantum projection noise \cite{ITA93}. On a Zeeman-shifted transition with a magnetic moment difference of 1 Bohr magneton ($\Delta \nu/\Delta B$ = 14 MHz/mT) when probed by the Ramsey method with free precession time $T_R$, the magnetic field resolution $\Delta B$ is \begin{equation}
\Delta B(\tau) = \frac{1}{2 \pi (\Delta \nu/\Delta B) \sqrt{T_R \tau}}=1.1\times
10^{-11} \rm{T}/\sqrt{\tau/\rm{s}},
\end{equation}
where $\tau$ is the averaging time and the last expression assumes $T_R$~= 1~s. By actively compensating the surface field with external coils that lead to a known field geometry, we could detect not only the modulus but also the direction of the local magnetic field. While magnetic fields can be sensed closer to the surface utilizing nitrogen vacancy centers \cite{BAL08, TAY08} and with higher signal-to-noise ratio with a large number of cold neutral atoms \cite{VEN07}, a potential advantage of the ion sensor could be the combined sensitivity to electric and magnetic fields.

In principle the electric field in all space can be reconstructed from the field in one plane, but this inversion is an ill-posed problem for spatial features much smaller than the ion-to-feature distance. Therefore the lateral spatial resolution when scanning the surface will be limited by the attained signal-to-noise ratio and is roughly equal to the distance to the surface if the signal-to-noise ratio is of order 1.

An interesting application of the ion sensor in quantum information processing with trapped ions is to use it for straightforward comparisons of heating rates of different surfaces \cite{DES06,TUR00,LAB08}. Typically, to compare heating rates from different electrode surfaces, separate traps composed of different materials have been built and tested. This leads to uncertainties due to variations in the trap geometry, the exact steps of materials processing, surface contamination due to cleaning agents, and the bake-out procedure. Most of these variables could be eliminated, and the testing of different materials could be accelerated by use of one ion sensor on a variety of material samples deposited on the same carrier surface (Fig. {\ref{fig:Apps}}b).

\begin{acknowledgments}
This work was supported by IARPA and the NIST Quantum Information Program.
We thank C. Ospelkaus and S. Ospelkaus for comments on the manuscript. This paper is a
contribution by the National Institute of Standards and Technology
and not subject to U.S. copyright.
\end{acknowledgments}


\bibliography{references}

\end{document}